\documentclass{article} 
\usepackage{iclr2020_conference,times}

\usepackage{amsmath,amsfonts,bm}

\def\eqref#1{equation~\ref{#1}}

\def\1{\bm{1}}

\DeclareMathAlphabet{\mathsfit}{\encodingdefault}{\sfdefault}{m}{sl}
\SetMathAlphabet{\mathsfit}{bold}{\encodingdefault}{\sfdefault}{bx}{n}

\usepackage{hyperref}
\usepackage{url}

\title{Politics of Adversarial Machine Learning}

\author{Kendra Albert
\thanks{Authors ordered alphabetically}\\
Harvard Law School\\
Cambridge, USA \\
\texttt{kalbert@law.harvard.edu} \\
\And
Jonathon Penney\\
Citizen Lab, University of Toronto\\
Toronto, Canada \\
\texttt{jon@citizenlab.ca} \\
\And
Bruce Schneier\\
Harvard Kennedy School\\
Cambridge, USA\\
\texttt{schneier@schneier.com}\\
\And
Ram Shankar Siva Kumar\\
Microsoft\\
Redmond, USA\\
\texttt{Ramk@Microsoft.com}
}

 \iclrfinalcopy 
\begin{document}

\maketitle

\begin{abstract}
 In addition to their security properties, adversarial machine-learning attacks and defenses have political dimensions. They enable or foreclose certain options for both the subjects of the machine learning systems and for those who deploy them, creating risks for civil liberties and human rights. In this paper, we draw on insights from science and technology studies, anthropology, and human rights literature, to inform how defenses against adversarial attacks can be used to suppress dissent and limit attempts to investigate machine learning systems, using facial recognition technology as a case study. To make this concrete, we use real-world examples of how attacks such as perturbation, model inversion, or membership inference can be used for socially desirable ends. Although this analysis' predictions may seem dire, there is hope. Efforts to  address human rights concerns in the commercial spyware industry provide guidance for similar measures to ensure ML systems serve democratic, not authoritarian ends.
\end{abstract}

\section{Introduction}
All technological work has some political dimension. As Langdon Winner illustrated 40 years ago, a technology's design, systems, or arrangements can pave the way for certain social or political relations or foreclose specific possibilities \citep{winner1980artifacts}.  Winner's most often cited example was the low bridges that crossed over the parkways that went between New York City and Long Island. Although it may seem that bridges with low clearance are not political, Winner explains that there was evidence suggesting that the underpasses were built this way to preclude public buses from using the roads --- denying  those who relied on public transit, predominantly low-income New Yorkers of color,  access to certain public spaces \citep{winner1980artifacts, woolgar1999artefacts}. Today, we can speak of the politics of algorithms, which can automate decisions in discriminatory ways \citep{noble2018algorithms, eubanks2018automating, benjamin2019race}  and spyware, which is used and abused by authoritarian governments to track, suppress, and harm human rights activists and dissidents \citep{harkin2019commodification, penney2018advancing}.  Technology has the potential to reinforce or undermine existing power relationships in the context in which it is used, and even technologies or related practices that  appear neutral, benign, or even benevolent  have potential impacts on civil liberties and human rights.

In cryptography, significant attention has been devoted to unpacking how particular research directions within the field may have implications for who can deploy cryptographic technologies \citep{rogaway2015moral}. There is also significant work that takes on these questions in the context of data science, Machine Learning (ML), and algorithmic decision-making more generally \citep{greendata, mac2018politics}. We turn a similar lens on the development of defenses against adversarial attacks on machine learning, exploring how efforts to secure machine learning systems against attacks can have real-world harms that disproportionately fall on those who wish to resist the use of such systems. Our conclusion is that those engaged in security work must understand that securing machine learning systems has consequences for the human rights and civil liberties of the subjects of those systems, consequences that Winner would describe as “political.”

In this paper, we first discuss how machine learning system deployments increase the chance that adversarial attacks will be used by the subjects of the systems, who are unlikely to have a full say in their construction or deployment. Second, we provide examples of how adversarial attacks could be used for desirable aims. Securing systems against attack may inadvertently suppress dissent, or foreclose research that aims to shed light on how ML systems harm particular populations. We also discuss how the “adversarial arms race” may lead to the development and deployment of more invasive forms of surveillance. Finally, we conclude by suggesting directions forward.  We draw an explicit connection to spyware, where activists, researchers, and civil society organizations have come together to recommend methods to reduce the harm of surveillance technologies. 

In this paper, we use facial recognition as a case study to demonstrate the politics of adversarial machine learning. Facial recognition technologies  (FRT) have been widely critiqued by scholars and activists \citep{garvie2016perpetual, malikaface}, and several municipalities within the United States have banned the use of FRT by local law enforcement \citep{mont2019face}. Most salient to our argument, the harms of FRT usage by authoritarian governments are not theoretical, and the scope and usages of FRT are expanding rapidly \citep{mozur2019one, balaban2015deep}. We note that the politics of securing machine learning systems extend beyond FRT discussed in this paper. 

\section{Political Implications of the Usage of ML and the Use of Adversarial Attacks}

The goals of implementing many machine learning systems can be summarized as making decisions “at scale,” generally without human intervention. Machine learning systems provide what science and technology studies author James C. Scott would call “legibility” \citep{scott1998seeing, thompson1967time}.  Scott defines legibility as the process by which states took “exceptionally complex, illegible, and local social practices, such as land tenure customs or naming customs, and created a standard grid whereby information could be centrally recorded and monitored.” A machine learning system that aims to tag an image with the items it contains is a literal example of legibility; it makes readable that which was previously not visible. 

Facial recognition technology makes that which only used to be done by humans (telling if a face matched) possible by machines. It is the combination of scale and legibility that makes machine learning systems uniquely attractive to governments and other institutions that seek to maintain control over large populations \citep{eubanks2018automating}. In the words of Meredith Whittaker, director of AI Now, “facial recognition is usually deployed by those who already have power, say employers, landlords, and the police --- to surveil, control, and in some cases oppress those who don't.” 
Current adversarial machine learning scholarship focuses on building robust ML systems and identifying novel attacks. Little attention has been paid to the people who are subjected to these ML systems. Subjects may not have the option of opting out or using democratic processes to control the systems, because of the speed and scale of deployment. As we were writing this paper, news broke of ClearviewAI, a company that scrapes publicly available internet images to develop an application that, at least theoretically, allows for the identification of any person's face. In order to opt out, you are required to send them an image of your government identification. In other cases, people are completely omitted. For instance, in 2019, DARPA, the research wing of the US Department of Defense, announced a challenge to build robust ML defenses, saying, “We must ensure ML is safe and incapable of being deceived” \citep{hava2019} --- but it never explicitly states safe from whom? And deceived by whom? Although it is common within the security community to view those who wish to interfere with the confidentiality, integrity, or availability of systems as “attackers,” this framing belies the fact that those who resist such systems could just as easily be pro-democracy protesters or academics interested in evaluating the inclusiveness of training data as they could be “malicious” actors.

\section{“Desirable” Attacks on ML}
To an ML system, an attacker motivated by a legitimate human rights and civil liberties concern with the system and an attacker motivated to hide something from the system are the same \citep{cowen2014deadly, maly2014inspection} A facial recognition system cannot tell if a person wearing a mask is a protester or a bank robber. Below, we discuss how three different adversarial attacks could be used for socially beneficial methods.

\begin{itemize}
\item  Membership inference can be used to determine whether a given data record was part of a model's training set or not. Hardening against membership inference could prevent a researcher from determining whether a given person was included, which can be useful for efforts at machine learning accountability or determining the source of images for dataset training. \cite{fredrikson2015model} use their membership inference techniques to determine whether a given picture was present in a facial recognition database. Determining whether a given image is present could help an individual determine whether they are able to bring a court case against a given facial recognition provider.
\item Adversarial examples \citep{goodfellow2014explaining} involve modification of a query to get a desired result. Defenses against perturbation attacks aim at hardening models against common modifications in order to still allow for image analysis. Its beneficial use was studied by \cite{kulynych2020pots} in the context of evading facial recognition systems. Similarly, the EqualAIs project runs a detection perturbation algorithm for the purpose of allowing individuals to make a certain image less likely to be detectable as a face \citep{equalai}. Obfuscation of this kind could be used by photographers who are documenting protests and would like to post the images without the potential for facial recognition software to automatically identify protesters \citep{blake2019hk}. 
\item Hardening against model inversion attacks \citep{fredrikson2015model} aim to prevent retrieval of private features. But when blackbox machine learning systems are deployed in contexts like access to credit, using attacks to retrieve the models may be one of the only ways to determine whether decisions are being made based on impermissible factors, such as race or gender.
\end{itemize}

\section{Adversarial Arms Race }
It's not just the preclusion of certain forms of attacks that has implications for the rights and liberties of people and groups that are subjects of machine learning systems. Efforts to secure ML systems against attack without proper attention being paid to the uses those systems may in fact lead to more invasive surveillance measures. \cite{biggio2018wild} remarked that securing ML systems is an “arms race” with attackers attempting to break into the system, and the defenders attempting to build robust defenses. 
Consider the following not-so-hypothetical scenario:
\begin{enumerate}
\item A surveillance state is using facial recognition to quash a peaceful protest. Dissidents try to escape facial recognition by using masks to occlude their faces. 
\item In order to defend against these attackers who are compromising the integrity of the FRT systems, the state steps up the game with structured occlusion coding for “robust” face recognition \citep{wen2016structured}; or using pre-trained model of full frontal faces to remove occlusion\citep{elmahmudi2019deep} In order to defend against this, the dissidents turn to using adversarial clothing to completely evade the FRT.
\item To gain the upper hand again, the surveillance state uses the newly released adversarial robustness toolkit from Baidu \citep{goodman2020advbox} that can help defend against adversarial clothing “attacks.” The dissidents now attempt to escape detection by wearing 3D-printed adversarial eyeglasses \citep{sharif2016accessorize}
\item To counter this, the surveillance state completely bypasses faces and uses other biometric technologies, such as iris scanning or gait detection \citep{hofmann2011gait}, to identify people. 
\end{enumerate}

This example may seem speculative and far-fetched (protesters wearing adversarial eyeglasses?) but they illustrate the way in which standard security arms-race thinking can lead to the deployment of more draconian surveillance measures that suppress dissent.

\section{Direction Forward}
Adversarial machine learning is not the only security technology with consequences for human rights. The commercial spyware industry also poses a similar threat, and efforts to address these risks can provide a useful guidance for those engaged in ML security and the development and commercialization of adversarial ML toolkits. Spyware researchers have extensively documented how the multimillion-dollar commercial spyware industry has been used by authoritarian governments around the world to track, suppress, and censor human rights activists and civil society groups \ \citep{harkin2019commodification, penney2018advancing}.  In response, a range of civil society, governmental, and industry actors have developed a range of ethical, corporate social responsibility, and human rights measures for commercial spyware industry participants \citep{antis2019, lauterbach2017no, access2019, mackune2019}  including industry standards for transparency, human rights due diligence, and commitments to "human rights by design" principles.
Similar proposals, using these recommendations as a foundation, could be offered for the ML industry, particularly those engaged in securing ML systems and developing and distributing ML toolkits.  We recommend the following action items:

\begin{itemize}
\item Vendors who sell machine learning systems should commit to the application of the UN Guiding Principles on Business and Human Rights (UN Guiding Principles) to the industry, and commit to multi-stakeholder efforts to establish and operationalize their requirements within the industry, particularly on transparency, human rights due diligence, and remedies. 

\item  Vendors should establish and comply with industry-wide standards for transparency and human rights policy/due diligence measures; blacklisting clients and customers based on human rights considerations (e.g., a government with a poor human rights record); and prohibitions on assisting clients with reconfiguring/hardening ML systems to resist attackers in contexts where human rights or civil liberties are at risk (e.g., protesters resisting FRT deployed by an authoritarian government). 

\item Both vendors and individual product teams should commit to human rights-by-design principles, whereby ML systems and toolkits would be designed to make abusive deployment more difficult. For instance, implementing mandatory features that report and log ML system uses as well as the nature and form of attacks; or automatic tool disablement on detection of misuse or attempts at reconfiguration. 

\item Practitioners should proactively ask questions about how the products that they secure are deployed and whether adequate safeguards, such as those discussed above, are in place to ensure that human rights are preserved in their creation, sale, and use. 

\end{itemize}

None of these ideas is perfect. Indeed, the lesson of attempts to combat authoritarian use of spyware has been that structured violation of human rights are difficult to overcome through voluntary practices. Nonetheless, these practices provide a framework for industry participants to deal with ML adversarial tools and practices.

\section{Conclusion}
Adversarial ML is at a pivotal moment. As these systems become more widely deployed, theoretical attacks and defenses rooted in the academic literature will become the stuff of people's lives. We have merely scratched the surface of what a political analysis of adversarial machine learning attacks and defenses might illuminate. The adversarial ML community has the opportunity to learn from scholars of science and technology studies, anthropology, and critical race theory --- as well as human rights and ethics literature more generally --- and to be in conversation with protesters, researchers, and others who seek to attack systems for socially beneficial reasons. Through understanding lived experiences of resistance, applying the lessons of other disciplines, as well as reflecting upon the work of those seeking to prevent similar outcomes with spyware, the adversarial ML community can not just understand its work as political but take affirmative steps to ensure that it is used primarily for good.

\subsubsection*{Acknowledgments}
We would like to thank Beth Friedman for her thoughtful feedback and edits to the paper. 

\appendix

\bibliography{iclr2020_conference}
\bibliographystyle{iclr2020_conference}

\end{document}